# Complexity to Find Wiener Index of Some Graphs

*Kalyani Das*

Department of Mathematics
Ramnagar College, Purba Medinipur, West Bengal, India

*Abstract.* The Wiener index is one of the oldest graph parameter which is used to study molecular-graph-based structure. This parameter was first proposed by Harold Wiener in 1947 to determining the boiling point of paraffin. The Wiener index of a molecular graph measures the compactness of the underlying molecule. This parameter is wide studied area for molecular chemistry. It is used to study the physio-chemical properties of the underlying organic compounds. The Wiener index of a connected graph is denoted by W(G) and is defined as $W(G) = \frac{1}{2}\sum_{u,v} d(u,v)$, that is W(G) is the sum of distances between all pairs (ordered) of vertices of G. In this paper, we give the algorithmic idea to find the Wiener index of some graphs, like cactus graphs and intersection graphs, viz. interval, circular-arc, permutation, trapezoid graphs.

## 1. Introduction

Molecular descriptor is a final result of a logic and mathematical procedure which transforms chemical information encoded with in a symbolic representation of a molecule into a useful number or the result of some standardized experiment. The Wiener index **W(G)** is a distance-based topological invariant is also a molecular descriptor, it much used in the study of the structure-property and the structure-activity relationships of various classes of biochemically interesting compounds introduced by Harold Wiener in 1947 for predicting boiling points ($b.p$) of alkanes based on the formula $b.p = \alpha W + \beta w(3) + \gamma$, where $\alpha, \beta, \gamma$ are empirical constants, and **w(3)** is called path number. It is defined as the half sum of the distances between all pairs of vertices of G[1,12,14]

$$W(G) = \frac{1}{2}\sum_{u,v} d(u,v)$$

## 2. Algorithms to find Wiener index
### 2.1. Cactus graphs

The class of cactus graph is an important subclass of general planar graphs. Let $G = (V, E)$ be a finite, connected, undirected simple graph of $n$ vertices $m$ edges, where $V$ is the set of vertices and $E$ is the set of edges.

A vertex $u$ is called a *cutvertex* if removal of $u$ and all edges incident on $u$ disconnect the graph. A connected graph without a cutvertex is called a *non-separable* graph. A *block* of a graph is a maximal non-separable subgraph. A *cycle* is a connected graph (or subgraph) in which every vertex is of degree two. A block which is a cycle is called a *cyclic block*. A *cactus graph* is a connected graph in which every block is either an edge or a cycle. A *weighted* graph $G$ is a graph in which every edge is associates with a weight. Without loss of generality we assume that all weights are positive. A *weighted cactus graph* is a weighted, connected graph in which every block containing two vertices is an edge and three or more vertices is a cycle. A *path* of a graph $G$ is an alternating sequence of distinct vertices and edges which begins and ends with vertices in $G$. The *length* of a path is the sum of the weights of the edges in the path. a path from vertex $u$ to $v$ is a *shortest path* if there is no other path from $u$ to $v$ with lower length. The *distance* $d(u,v)$ between vertices $u$ and $v$ is the length of shortest path between $u$ and $v$ in $G$.

**Theorem 1.** [13] *The shortest distances from a specified vertex to all other vertices of a weighted cactus graph can be computed in $O(n)$ time and the all pair shortest distance of a weighted cactus graph can be computed in $O(n^2)$ time, where $n$ represents the total number of vertices of the graph.*

**Theorem 2.** *The Winner index of the cactus graphs can be computed in $O(n^2)$ time, where n represents the total number of vertices of the graph.*

**Definition of Intersection graphs**
A graph $G = (V, E)$ is called an *intersection graph* for a finite family $F$ of a non-empty set if there is a one-to-one correspondence between $F$ and $V$ such that two sets in $F$ have non-empty intersection if and only if their corresponding vertices in $V$ are adjacent. We call $F$ an intersection model of $G$. For an intersection model $F$, we use $G(F)$ to denote the intersection graph for $F$.

Depending on the nature or geometric configuration of the sets $S_1, S_2,...$ different types of intersection graphs are generated. The most useful intersection graphs are
- Interval graphs ($S$ is the set of intervals on a real line)
- Tolerance graphs
- Circular-arc graphs ($S$ is the set of arcs on a circle)
- Permutation graphs ($S$ is the set of line segments between two line segments)
- Trapezoid graphs ($S$ is the set of trapeziums between two line segments)
- Disk graphs ($S$ is the set of circles on a plane)
- Circle graphs ($S$ is the set of chords within a circle)
- Chordal graphs ($S$ is the set of connected subgraphs of a tree)
- String graphs ($S$ is the set of curves in a plane)

- Graphs with boxicity $k$ ($S$ is the set of boxes of dimension $k$)
- Line graphs ($S$ is the set of edges of a graph).

### 2.2. Interval graphs

An undirected graph $G = (V, E)$ is said to be an *interval graph* if the vertex set $V$ can be put into one-to-one correspondence with a set $I$ of intervals on the real line such that two vertices are adjacent in $G$ if and only if their corresponding intervals have non-empty intersection. That is, there is a bijective mapping $f : V \to I$.

The set $I$ is called an interval representation of $G$ and $G$ is referred to as the interval graph of $I$ [19]. A large number of work on intersection graphs and cactus graphs have been done in [20-37].

**Theorem 3.** *[7] The time complexity for finding the distances between all pair of vertices on interval graphs is $O(n^2)$.*

**Theorem 4.** *The Winner index of the interval graphs can be computed in $O(n^2)$ time, where n represents the total number of vertices of the graph.*

### 2.3. Circular-arc graphs

A graph is a *circular-arc* graph if there exists a family $A$ of arcs around a circle and a one-to-one correspondence between vertices of $G$ and arcs in $A$, such that two distinct vertices are adjacent in $G$ if and only if the corresponding arcs intersect in $A$. Such a family of arcs is called an *arc representation* for $G$.

A graph $G$ is a *proper circular-arc (PCA)* graph if there exists an arc representation for $G$ such that no arc is properly included in another.

A graph $G$ is a *unit circular-arc (UCA)* graph if there exists an arc representation for $G$ such that all the arcs are of the same length.

**Theorem 5.** *[18] The all-pair shortest paths problem on circular-arc graph is computed in $O(n^2)$ time.*

**Theorem 6.** *The Winner index of the circular-arc graph is computed in $O(n^2)$ time.*

### 2.4. Permutation graphs

An undirected graph $G = (V, E)$ with vertices $V = \{1, 2, \ldots, n\}$ is called a permutation graph if there exists a permutation $\pi$ on $N = \{1, 2, \ldots, n\}$ such that for all $i, j \in N$,

$$(i - j)(\pi^{-1}(i) - \pi^{-1}(j)) < 0$$

if and only if $i$ and $j$ are joined by an edge in $G$ [19]. Geometrically, the integers $1,2,\ldots,n$ are drawn in order on a real line called as *upper line* and $\pi(1),\pi(2),\ldots,\pi(n)$ on a line parallel to this line called as *lower line* such that for each $i \in N$, $i$ is directly below $\pi(i)$. Next, for each $i \in V$, a line segment is drawn from $i$ on the lower line to $i$ on the upper line and it is denoted by $l(i)$. Then from definition it follows that there is an edge $(i, j)$ in $G$ if and only if the line segment $l(i)$ for $i$ intersects the line segment $l(j)$ for $j$.

**Theorem 7.** *[17] The all-pair shortest paths problem on permutation graphs in $O(n^2)$ time.*

**Theorem 8.** *The Winner index of the permutation graphs can be computed in $O(n^2)$ time.*

### 2.5. Trapezoid graphs

A trapezoid $T_i$ is defined by four corner points $[a_i, b_i, c_i, d_i]$, where $a_i < b_i$ and $c_i < d_i$ with $a_i, b_i$ lying on top line and $c_i, d_i$ lying on bottom line of a rectangular channel. An undirected graph $G = (V, E)$ with vertex set $V = \{v_1, v_2, \ldots, v_n\}$ and edge set $E = \{e_1, e_2, \ldots, e_m\}$ is called a trapezoid graph if a trapezoid representation can be obtained such that each vertex $v_i$ in $V$ corresponds to a trapezoid $T_i$ and $(v_i, v_j) \in E$ if and only if the trapezoids $T_i$ and $T_j$ corresponding to the vertices $v_i$ and $v_j$ intersect. For simplicity the vertices $v_1, v_2, \ldots, v_n$ are represented respectively by 1, 2, ..., $n$. Thus the edge $(i, j) \in E$ if and only if $T_i$ and $T_j$ intersect in the trapezoid representation.

**Theorem 9.** *The time complexity to find all pairs shortest distances on trapezoid graphs is $O(n^2)$.*

**Theorem 10.** *The time complexity to compute Winner index of a trapezoid grapg is $O(n^2)$.*

Wiener Index of a Cycle in the Context of Some Graph Operations